\documentclass{PoS}

\def\si{^1 \hskip -0.02in S _0}
\def\siii{^3 \hskip -0.02in S _1}

\title{Hadronic interactions and nuclear physics }

\ShortTitle{Hadronic interactions and nuclear physics }

\author{\speaker{Silas R.~Beane}\\
        Department of Physics, University of New Hampshire, Durham NH 03824 \\
        E-mail: \email{silas@physics.unh.edu}}

\abstract{I give an overview of efforts in the last year to calculate
interactions among hadrons using lattice QCD. Results discussed
include the extraction of low-energy phase shifts and three-body
interactions, and the study of pion and kaon condensation. A critical
appraisal is offered of recent attempts to calculate nucleon-nucleon
and nucleon-hyperon potentials on the lattice.}

\FullConference{The XXVI International Symposium on Lattice Field Theory \\
                 July 14 - 19, 2008\\
                 Williamsburg, Virginia, USA}

\begin{document}

\section{Motivation}

\noindent Lattice QCD promises to revolutionize our understanding of
multi-hadron systems and nuclear physics. Being one level of
difficulty removed from the physics of single hadrons, the study of
nuclei has traditionally been confined to phenomenological
descriptions which are, for the most part, disconnected from the
Standard Model of particle physics. Over the past decade, a great deal
of progress has been made in formulating various low-energy effective
field theories (EFT) of QCD which describe few-nucleon
systems~\cite{Bedaque:2002mn,Epelbaum:2005pn}, and even finite
nuclei~\cite{Navratil:2007we}. The study of many-hadron systems using
lattice QCD has only begun very recently, with accurate results now
available for some meson systems, as we will see in this review.

The underlying motivation for studying complicated hadronic systems
(like nuclei) varies. While many lattice theorists are interested in
calculating hadronic quantities that are relevant for an understanding
of physics beyond the standard model, there exists an entirely
independent motivation: there are many intrinsically interesting
hadronic quantities for which there is essentially no experimental
information, and for these quantities lattice QCD calculations will
have substantial impact. For instance, while there were several
experiments decades ago which measured the low-energy hyperon-nucleon (YN)
interaction, very little is known about such basic scattering
quantities as the scattering lengths and effective ranges\footnote{
There is some movement to remedy this deficiency; for instance, a
recent YN scattering experiment at relatively high
energies is described in Ref.~\cite{Kanda:2006bz}.}  And yet, for
instance, the $\Sigma^- n$ interaction is an important ingredient in
the nuclear Equation of State that determines the fate of dense
astrophysical objects like neutron stars~\cite{Beane:2008dv}.  Yet
another quantity of interest is $h_{\pi NN}$, the parity-violating,
flavor conserving, pion-nucleon coupling constant.  After many decades
of intense experimental effort this quantity remains mysterious, and
while a lattice determination poses serious
challenges~\cite{Beane:2002ca}, the required computer-time resources
constitute a minute fraction of what is required for an experiment in
nuclear or particle physics.  As a final example, the $K\pi$
scattering lengths have recently been determined using lattice
QCD~\cite{Beane:2008dv}, providing a prediction for the global
experimental effort led by the DIRAC collaboration~\cite{DIRACprops}
to measure these quantities by observing the decays of mesonic atoms.

As is well known, lattice correlation functions involving baryons face
a significant signal to noise problem; i.e. signal to noise degrades
exponentially with time~\cite{Lepage:1989hd}. Since this issue is so
fundamental to lattice QCD studies of nuclear physics, I will first
review this basic result. I will then discuss progress over the last
year in calculating hadron-hadron and multi-hadron interactions using
lattice QCD.  Before calculating a scattering process that is unknown
or poorly known experimentally, it is essential to ``benchmark''
against quantities that are well known either from experiment and/or
from independent theoretical considerations. The s-wave $\pi\pi$
scattering lengths offer a powerful means of benchmarking lattice QCD
methods as these quantities are known with remarkable accuracy from
the Roy equation method~\cite{Colangelo:2001df}. I will discuss a
recent lattice QCD calculation of the $I=2$ $\pi\pi$ scattering length
which has achieved accuracy at the $1\%$ level. Other recent results
in the meson sector will also be mentioned, including $K^+K^+$
scattering and the interactions of up to twelve pions and kaons, which
allow determinations of three-body interactions as well as chemical
potentials relevant to a description of pion and kaon condensation.

In meson-baryon scattering, benchmarking is significantly more
difficult because the channels which have been calculated to date do
not have disconnected diagrams and these are not well known
experimentally. I will review the status of these calculations.

In nucleon-nucleon scattering, benchmarking the s-wave scattering
length poses an extreme challenge to lattice QCD theorists. On the one
hand, there is a severe signal to noise problem which requires vast
computer resources to overcome. Then there is the problem of
extrapolation: since the physical scattering lengths are fine tuned,
the chiral EFT description is necessarily
non-perturbative. Furthermore, the radius of convergence of the EFT is
smaller than in the sector with mesons or a single baryon, and
therefore smaller quark masses are required in order to extrapolate.
In spite of these difficulties, a great deal of progress is being made
in providing a lattice postdiction of the threshold s-wave NN
scattering parameters. While a fully-dynamical lattice QCD calculation
of low-energy YN phase shifts has been carried out, presently one can
only compare to model calculations whose reliability is not clear.
I will review existing results for NN and YN scattering.

Recent calculations of NN and YN potentials on the lattice have
generated a great deal of interest. Unfortunately, the flaws in these
calculations have not engendered as much attention.  Therefore, I will
give a detailed critique of recent attempts to calculate NN and YN
potentials using lattice QCD.

\section{Signal to Noise Estimates}

%%%%%%%%%%%%%%%%%%%%%%%%%%%%%%%%%%%%%%%%%%%%%%%%%%%%%%%%%%%%%%%%%%%%%%%%%%%%
\begin{figure}[hb!]
\begin{center}
\centerline{\includegraphics*[width=0.63\textwidth,clip,angle=0]{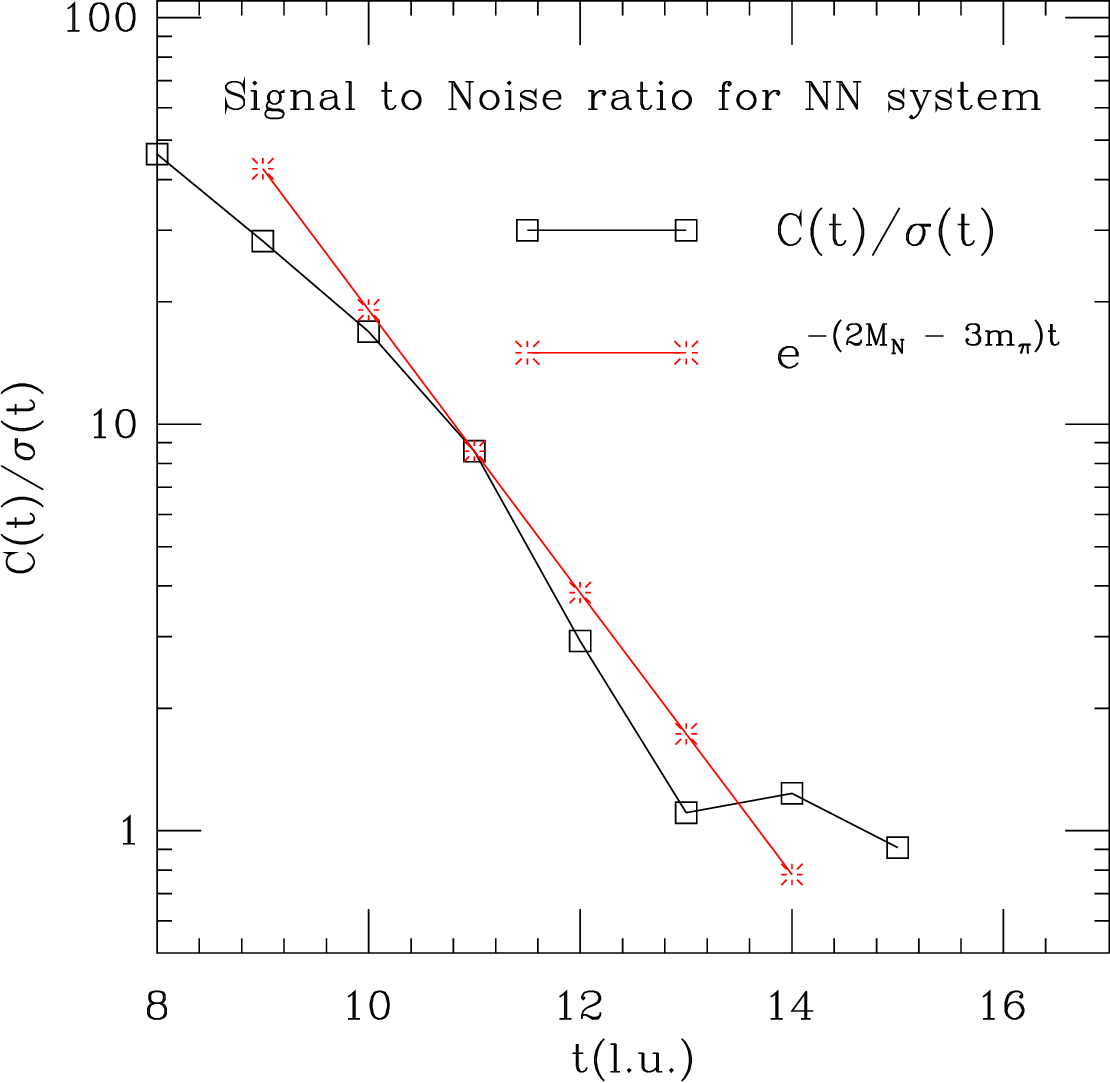}}
\caption{Signal to noise ratio for the NN system in the $\si$ channel. The lattice
data (black line) is generated from the MILC coarse ensemble with pion mass  $\sim350~{\rm MeV}$, 
as discussed in Ref.~\protect\cite{Beane:2008dv}. The theoretical prediction (red line)
is from eq.~\protect\ref{eq:NtoSnucleus}.} \label{fig:noise}
\end{center}
\end{figure}
%%%%%%%%%%%%%%%%%%%%%%%%%%%%%%%%%%%%%%%%%%%%%%%%%%%%%%%%%%%%%%%%%%%%%%%%%%%%%

\noindent As is well known~\cite{Lepage:1989hd}, very general
field-theoretic arguments allow a robust estimate of the noise to
signal ratio of hadronic correlation functions calculated on the
lattice. With an eye towards lattice QCD attempts to describe nuclei,
it is worth briefly noting the fundamental difference between
lattice-measured correlation functions involving mesons and
baryons. As an example, consider the noise to signal ratio of a
correlation function involving $n$ pion fields, where the small
interaction is neglected,
\begin{eqnarray}
{\sigma (t)\over \langle \theta(t) \rangle }&&
\sim \ {\sqrt{\left(A_2 - A_0^2 \right)} \ e^{-n m_\pi t}\over \sqrt{N} A_0\
e^{-n m_\pi t} }
\sim {1\over \sqrt{N}}
 \ \ .
\label{eq:NtoSpi}
\end{eqnarray}
Here $\langle \theta(t)\rangle$ is the correlation function, $\sigma
(t)$ is the variance and the $A_i$ are amplitudes. It is noteworthy
that in this ratio, the time dependence of the variance mirrors the
time dependence of the correlator itself. One therefore concludes that
correlators involving arbitrary numbers of pions have time-independent
errors, as is indeed observed in lattice calculations. This of course
renders the study of mesonic correlators quite pleasurable (from the 
statistical perspective).

The baryons provide a more disturbing story; consider the noise to signal
ratio for a proton correlation function:
\begin{eqnarray}
{\sigma (t)\over \langle \theta(t) \rangle }&&
\sim \ {\sqrt{A_2} \ e^{-{3\over 2} m_\pi t}\over \sqrt{N} A_0 e^{- m_p t} }
\sim   {1\over \sqrt{N}} \ e^{\left( m_p - {3\over 2} m_\pi\right) t} 
 \ \ .
\label{eq:NtoSproton}
\end{eqnarray}
Here the variance is dominated by the three-pion state rather than by
the proton, and therefore the noise to signal ratio of the proton
correlator grows exponentially with time.  More generally, for a
system of $A$ nucleons, the noise to signal ratio behaves as
\begin{eqnarray}
{\sigma (t)\over \langle \theta(t) \rangle }&&
\sim   {1\over \sqrt{N}} \ e^{\ A \left( m_p - {3\over 2} m_\pi\right) t}
 \ \ .
\label{eq:NtoSnucleus}
\end{eqnarray}
Therefore the situation worsens as one adds nucleons. Fig.~\ref{fig:noise}
compares the theoretical expectation from eq.~\ref{eq:NtoSnucleus} with
$A=2$ to data calculated by the NPLQCD collaboration~\cite{Beane:2008dv}.

These estimates, which follow from very general field theoretic
arguments, indicate that nucleon and nuclear physics require {\it
exponentially more resources} than meson physics to achieve the same
level of accuracy.  There has been an attempt to get around this
problem~\cite{Bedaque:2007pe} by eliminating the pion zero modes
through a clever choice of boundary conditions. However, this method
does not yet have a practical implementation. Clearly more 
effort should be dedicated to this very important problem.

\section{$n$ bosons in a box}

\noindent The ground-state energy of an $n$-boson system~\cite{Beane:2007qr} in a finite volume is
calculated with an interaction of the form
\begin{eqnarray}
  \label{eq:4}
  V({\bf r}_1,\ldots,{\bf r_n}) =
\eta \sum_{i< j}^n \delta^{(3)}({\bf r}_i-{\bf r}_j)
+\eta_3 \sum_{i< j<k}^n \delta^{(3)}({\bf r}_i-{\bf
  r}_k)\delta^{(3)}({\bf r}_j-{\bf r}_k) +..,
\end{eqnarray}
where $\eta$ and $\eta_3$ are the two- and three-body
pseudo-potentials, respectively, and the ellipsis denote higher-body
interactions. In general, $m$-body interactions will enter at ${\cal
O}(L^{3(1-m)})$ in the large volume expansion.  For an $s$-wave
scattering phase shift, $\delta(p)$, the two-body contribution to the
pseudo-potential is given by $\eta=-\frac{4\pi}{ M}
p^{-1}\tan\delta(p)= \frac{4\pi}{M}a + \frac{2\pi }{M}a^2 r p^2
+\ldots$, keeping only the contributions from the scattering length
and effective range, $a$ and $r$, respectively. To ${\cal O}(L^{-6})$
the coefficient of the three-body pseudo-potential, $\eta_3$, is
momentum independent.

As an example, consider the $2$-boson energy. The volume dependence of
the energy may be built up using Rayleigh-Schr\"odinger
time-independent perturbation theory. The leading contribution to the
perturbative expansion of the energy is given by
\begin{eqnarray}
\label{eq:a}
  {\Delta E}_2^{(1)} & = & \langle{-{\bf k},{\bf k}}| V({\bf r}_1,{\bf r}_2) |{-{\bf p},{\bf p}}\rangle \ ,
\end{eqnarray}
where $|{-{\bf p},{\bf p}}\rangle$ are the two-body momentum eigenstates
in the center-of-mass system. The single-particle wavefunctions in the
finite volume are given by: $\langle {\bf r}|{\bf p}\rangle = \exp(i
{\bf k}\cdot{\bf r})/L^{3/2}$.  Inserting two complete sets of
position eigenstates in eq.~(\ref{eq:a}), one finds 
\begin{eqnarray}
\label{eq:1}
{\Delta E}_2^{(1)}&=& \frac{\eta}{L^3}\ =\  \frac{4\pi\, a}{M\,L^3}\ .
\end{eqnarray}
One sees that in the large volume limit, the two-particle ground-state energy,
say, calculated on the lattice, is related to the scattering length.
(Excited levels then allow reconstruction of the entire phase shift.)
It is straightforward to calculate higher-order $1/L$ corrections in
this manner. Similarly, for $n$ bosons in a finite volume, one finds
\begin{eqnarray}
\label{eq:2}
{\Delta E}_n^{(1)}&=&  \left( \begin{array}{c} n \\ 2 \end{array} \right)
\frac{4\pi\, a}{M\,L^3}\ .
\end{eqnarray}
The volume dependence of the energy of the $n$-boson ground state in a
periodic cubic spatial volume of periodicity $L$ has now been
calculated~\cite{Huang:1957im,Lee:1957,Wu:1959,Luscher:1986pf,Luscher:1990ux,Beane:2007qr,Tan:2007bg,Detmold:2008gh}
up to ${\cal O}\left(1/L^{7}\right)$. While the calculational
framework described here is non-relativistic, the results remain valid
relativistically. In the two-body case, this has been shown by
L\"uscher~\cite{Luscher:1990ux}.  With $n\geq 3$ the interaction of
three particles due to the two-body interaction first enters at
$L^{-5}$, and relativistic effects in such interactions are suppressed
by $(ML)^{-2}$. Hence, the first relativistic effects occur at ${\cal
O}(L^{-7})$~\cite{Beane:2007qr} and have been calculated perturbatively in
Ref.~\cite{Detmold:2008gh}. The finite-volume energy formulas for fermions
are considered in Ref.~\cite{Luu:2008fg}.

\section{Meson-meson interactions}

%%%%%%%%%%%%%%%%%%%%%%%%%%%%%%%%%%%%%%%%%%%%%%%%%%%%%%%%%%%%%%%%%%%%%%%%%%%%
\begin{figure}[hb!]
\begin{center}
\centerline{\includegraphics*[width=0.4\textwidth,clip,angle=-90]{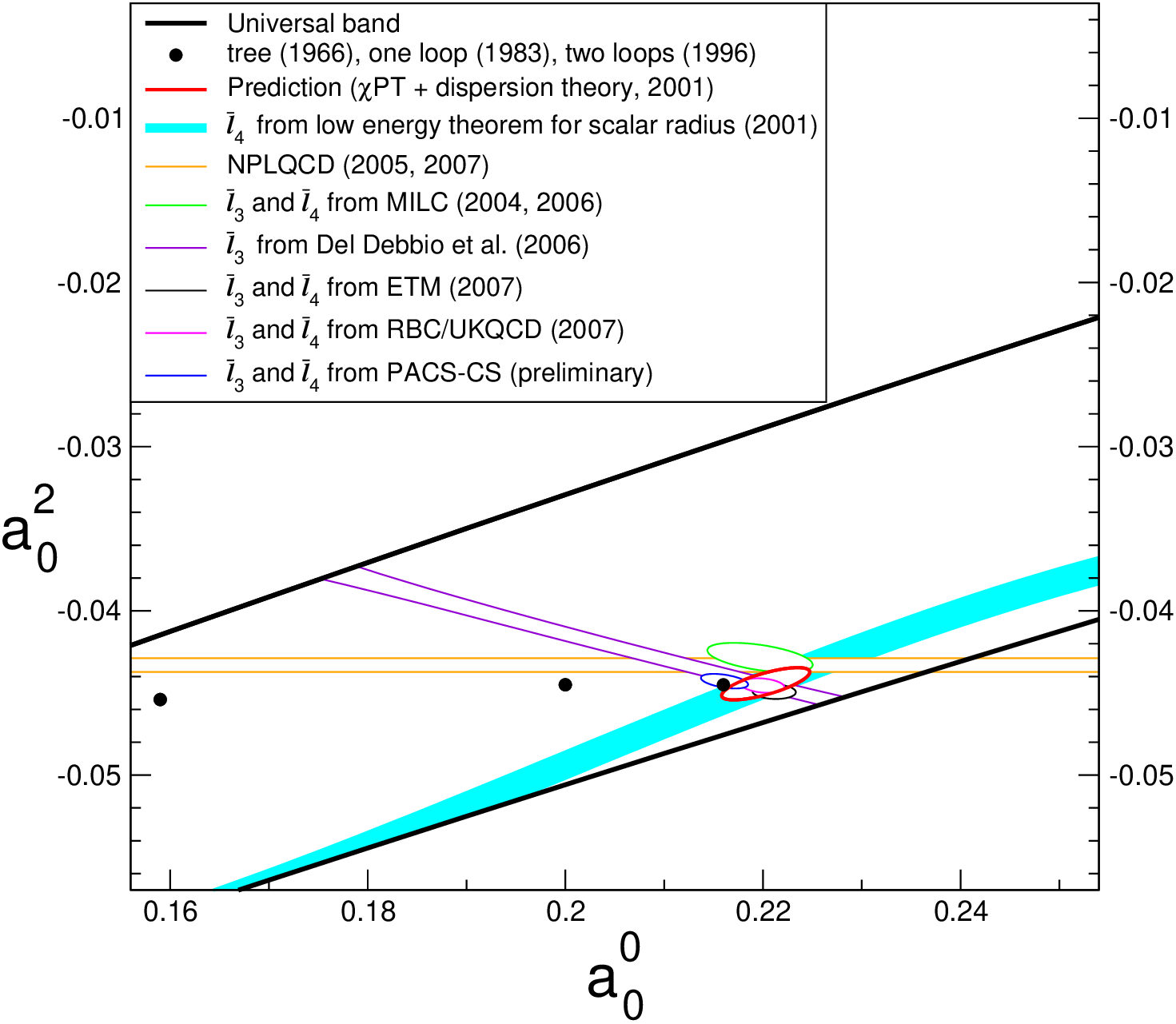}\includegraphics*[width=0.4\textwidth,angle=-90]{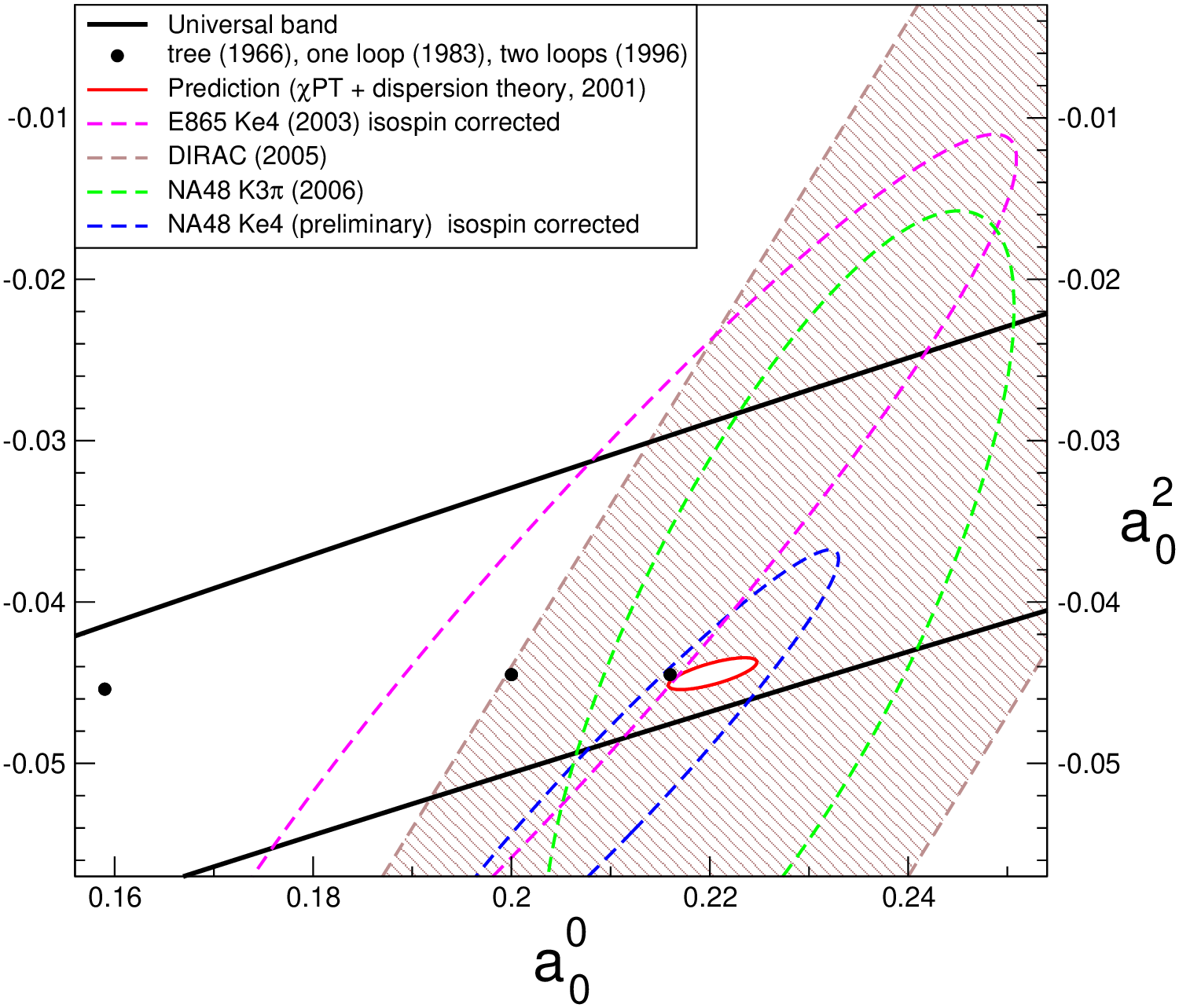}}
%\centerline{\psfig{file=a0a2_theory.ps,width=4.8cm,angle=-90}\psfig{file=a0a2_exp.ps,width=4.8cm,angle=-90}}
\caption{The state of threshold s-wave $\pi\pi$ scattering. Left panel: theoretical results. Noteworthy are
the red ellipse from the Roy equation analysis and the orange band from the lattice QCD calculation of
the $I=2$ scattering length, as discussed in the text. Right panel: experimental results. For detailed
information about all of the curves on these plot, see Ref.~\protect\cite{Leutwyler:2006qq}} \label{fig:pipi2}
\end{center}
\end{figure}
%%%%%%%%%%%%%%%%%%%%%%%%%%%%%%%%%%%%%%%%%%%%%%%%%%%%%%%%%%%%%%%%%%%%%%%%%%%%%

\noindent As the simplest application of eq.~(\ref{eq:1}), consider
recent results for the $\pi\pi$ interaction~\cite{Beane:2007xs}.  Due
to the chiral symmetry of QCD, pion-pion ($\pi\pi$) scattering at low
energies is the simplest and best-understood hadron-hadron scattering
process.  The scattering lengths for $\pi\pi$ scattering in the s-wave
are uniquely predicted at leading order in chiral perturbation
theory ($\chi$-PT)~\cite{Weinberg:1966kf}:
\begin{eqnarray}
m_\pi a_{\pi\pi}^{I=0} \ = \ 0.1588 \ \ ; \ \ m_\pi a_{\pi\pi}^{I=2} \ = \
-0.04537 
\ \ \ ,
\label{eq:CA}
\end{eqnarray}
at the charged pion mass. While experiments do not provide stringent
constraints on the scattering lengths, a determination of s-wave
$\pi\pi$ scattering lengths using the Roy equations has reached a
remarkable level of
precision~\cite{Colangelo:2001df,Leutwyler:2006qq}:
\begin{eqnarray}
m_\pi a_{\pi\pi}^{I=0} \ = \ 0.220\pm 0.005 \ \ ; \ \ m_\pi a_{\pi\pi}^{I=2} \ = \ -0.0444\pm 0.0010
\ \ \ .
\label{eq:roy}
\end{eqnarray}
The Roy equations~\cite{Roy:1971tc,Basdevant:1973ru,Ananthanarayan:2000ht},
use dispersion theory to relate scattering data at high energies
to the scattering amplitude near threshold. At present lattice QCD can
compute $\pi\pi$ scattering only in the $I=2$ channel as the $I=0$
channel contains disconnected diagrams. It is of course of great
interest to compare the precise Roy equation predictions with lattice
QCD calculations. Fig.~\ref{fig:pipi2} summarizes the theoretical
(left panel) and experimental (right panel) constraints on the s-wave
$\pi\pi$ scattering lengths~\cite{Leutwyler:2006qq}. It is clearly a strong-interaction
process where theory has outpaced the very-challenging experimental
measurements.

%%%%%%%%%%%%%%%%%%%%%%%%%%%%%%%%%%%%%%%%%%%%%%%%%%%%%%%%%%%%%%%%%%%%%%%%%%%%
\begin{figure}[hb!]
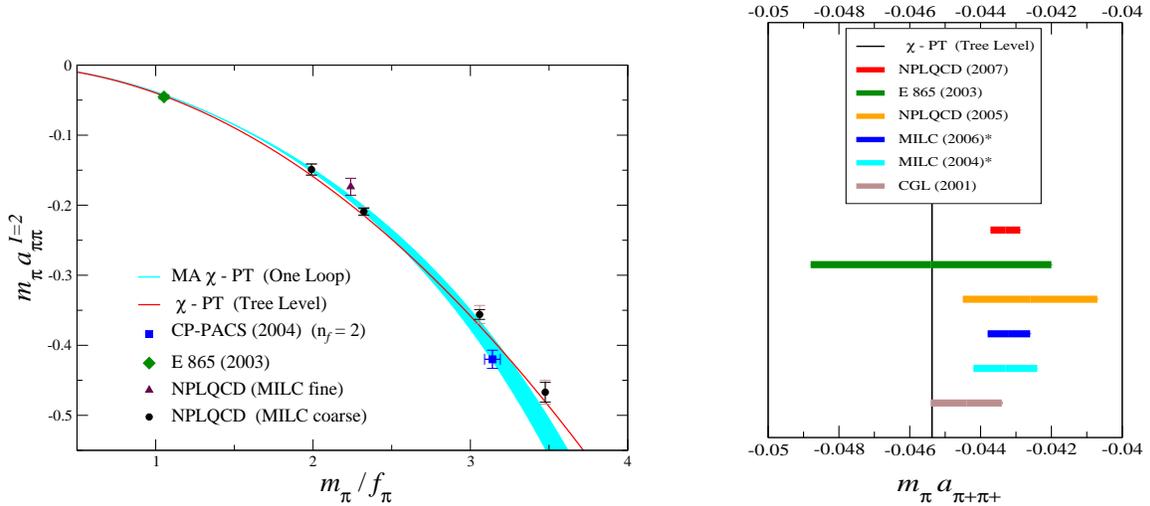

\begin{center}
\centerline{\includegraphics*[width=0.55\textwidth,clip]{mpia2Plot3_pipi.eps}\  \ \hfill\includegraphics*[width=0.35\textwidth,height=0.44\textwidth]{barchart2.eps}}
%\includegraphics[width=6 cm]{mpia2Plot3_pipi.eps}
%\hfill
%\includegraphics[width=5.5 cm]{barchart2.eps}
\caption{ Left panel: $m_\pi \ a_{\pi\pi}^{I=2}$ vs. $m_\pi/f_\pi$
(ovals) with statistical (dark bars) and systematic (light bars)
uncertainties.  Also shown are the experimental value from
Ref.~\protect\cite{Pislak:2003sv} (diamond) and the lowest quark mass result
of the $n_f=2$ dynamical calculation of CP-PACS~\protect\cite{Yamazaki:2004qb}
(square).  The blue band corresponds to a weighted fit to the lightest
three data points using the one-loop MA$\chi$-PT formula
(the shaded region corresponds only to the statistical error). The red line is the tree-level $\chi$-PT
result. Right panel: A compilation of the various measurements and
predictions for the $I=2$ $\pi\pi$ scattering length.  The prediction
described in these proceedings is labeled NPLQCD (2007), and the Roy equation determination of
Ref.~\protect\cite{Colangelo:2001df}  is labeled CGL (2001).} \label{fig:pipi}
\end{center}
\end{figure}
%%%%%%%%%%%%%%%%%%%%%%%%%%%%%%%%%%%%%%%%%%%%%%%%%%%%%%%%%%%%%%%%%%%%%%%%%%%%%

The only existing fully-dynamical lattice QCD prediction of the $I=2$
$\pi\pi$ scattering length involves a mixed-action lattice QCD scheme
of domain-wall valence quarks on a rooted staggered sea. Details of
the lattice calculation can be found in
Refs.~\cite{Beane:2007xs,Savage:2008mh}.  The energy difference $\Delta
E_2$ (and via eq.~(\ref{eq:1}) the scattering length) was computed at
pion masses, $m_\pi\sim 290~{\rm MeV}$, $350~{\rm MeV}$, $490~{\rm
MeV}$ and $590~{\rm MeV}$, and at a single lattice spacing, $b\sim
0.125~{\rm fm}$ and lattice size $L\sim 2.5~{\rm
fm}$~\cite{Beane:2007xs}. The physical value of the scattering length
was obtained using two-flavor mixed-action $\chi$-PT (MA$\chi$-PT )
which includes the effect of finite lattice-spacing artifacts to
$\mathcal{O}(m_\pi^2 b^2)$ and $\mathcal{O}(b^4)$~\cite{Chen:2006wf}.
Figure~\ref{fig:pipi} (left panel) is a plot of $m_\pi \
a_{\pi\pi}^{I=2}$ vs. $m_\pi/f_\pi$ with the lattice results and the
fit curves from MA$\chi$-PT.  The final result is:
\begin{eqnarray}
m_\pi a_{\pi\pi}^{I=2} & = &  -0.04330 \pm 0.00042
\ \ \ ,
\label{eq:nplqcd2}
\end{eqnarray}
where the statistical and systematic uncertainties have been combined
in quadrature.  Notice that $1\%$ precision is claimed in this
result. This result is consistent with all previous determinations
within uncertainties (see Figure~\ref{fig:pipi} (right panel)).  In
particular the agreement between this result and the Roy equation
determination is a striking confirmation of the lattice methodology,
and a powerful demonstration of the constraining power of chiral
symmetry in the meson sector.

It would be of great interest to see other (fully-dynamical) lattice
QCD calculations of the s-wave $\pi\pi$ scattering lengths using
different types of fermions. Recently, $\chi$-PT for Wilson-type
quarks has been developed for $\pi\pi$
scattering~\cite{Buchoff:2008ve,Aoki:2008gy} with an eye towards
lattice calculations. One may also wonder about new methodologies for
computing scattering which compete with the finite volume
method. In this respect, there has been promising recent work on
calculating the phase shift from the two-pion
wavefunction~\cite{Sasaki:2008sv}.

The $I=3/2$, $K^+K^+$ scattering length has also been computed by the
NPLQCD collaboration~\cite{Beane:2007uh,Savage:2008mh}.  At the
physical value of $m_{K^+}/f_{K^+}$,
\begin{equation}
        m_{K^+}\  a_{K^+ K^+} \ =\  -0.352 \pm 0.016 \ ,
\end{equation}
where statistical and systematic errors have been added in quadrature.
This scattering parameter, which is not measured experimentally, may
be useful for the study of kaon interferometry in heavy-ion
collisions.

\section{Multi-Meson Interactions}

%%%%%%%%%%%%%%%%%%%%%%%%%%%%%%%%%%%%%%%%%%%%%%%%%%%%%%%%%%%%%%%%%%%%%%%%%%%%
\begin{figure}[hb!]
\begin{center}
\centerline{\includegraphics*[width=0.99\textwidth,clip,angle=0]{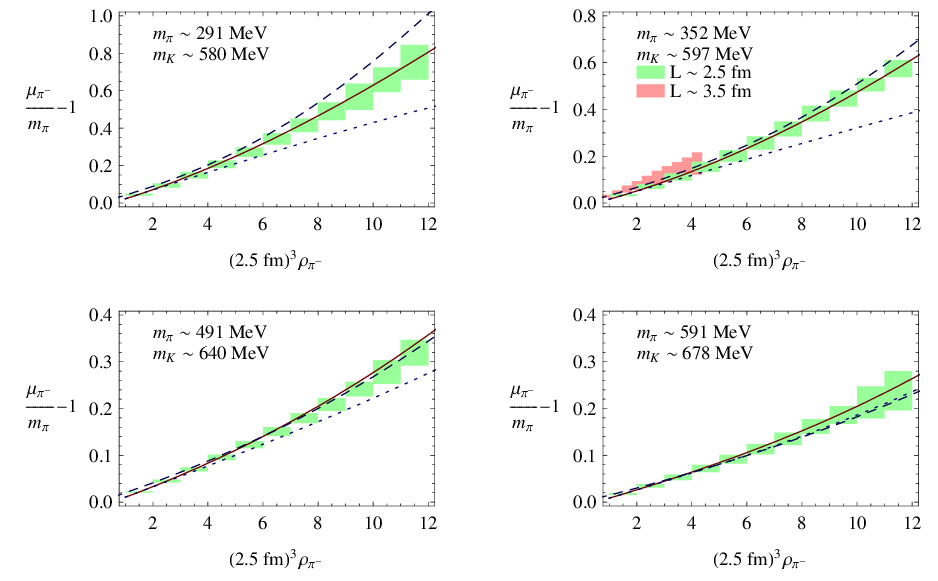}}
\caption{$K^-$ chemical potential as a function of $K^-$
density on the coarse MILC lattices. Finite differences obtained from lattice data appear as
boxes. The curves are: leading-order $\chi$-PT (dashed), fitted
scattering length with three-body interaction (solid) and same with no
three-body interaction (dotted)}
\label{fig:threebodforce}
\end{center}
\end{figure}
%%%%%%%%%%%%%%%%%%%%%%%%%%%%%%%%%%%%%%%%%%%%%%%%%%%%%%%%%%%%%%%%%%%%%%%%%%%%%

\noindent Perhaps surprisingly, lattice QCD calculations with up to
twelve pions and kaons have recently been carried out. Using the
large-volume expansion of the ground state energies of these systems,
it has proved possible to extract a signature of a three-pion
force~\cite{Beane:2007es,Detmold:2008fn,Detmold:2008yn,Detmold:2008bp}.
The difficulty with this result is that, unlike the relation between
two-body interactions and scattering, it is not obvious how to relate the
three-body interaction (which is proportional to the coefficient of a
three-body operator in a non-relativistic Lagrangian) to an observable
quantity. Nevertheless, Ref.~\cite{Detmold:2008fn} has noted that a
comparison can be made between $\chi$-PT predictions for pion
condensation~\cite{Son:2000xc} and the lattice QCD results. And indeed the
lattice-extracted three-body force appears to be essential for
agreement with the $\chi$-PT result, which in principle contains all
$n$-pion forces. Similar results have been found for kaons in
Ref.~\cite{Detmold:2008bp}; however the kaon three-body is consistent
with zero (See Fig.~\ref{fig:threebodforce}.)  This agreement is quite
remarkable given that the lattice calculation is clearly not in the
thermodynamic limit. This result demonstrates that lattice QCD
calculations with a finite number of particles are useful for the
study of many-body physics, like pion and kaon condensation.

%%%%%%%%%%%%%%%%%%%%%%%%%%%%%%%%%%%%%%%%%%%%%%%%%%%%%%%%%%%%%%%%%%%%%%%%%%%%
\begin{figure}[hb!]
\begin{center}
\centerline{\includegraphics*[width=1.1\textwidth,clip,angle=0]{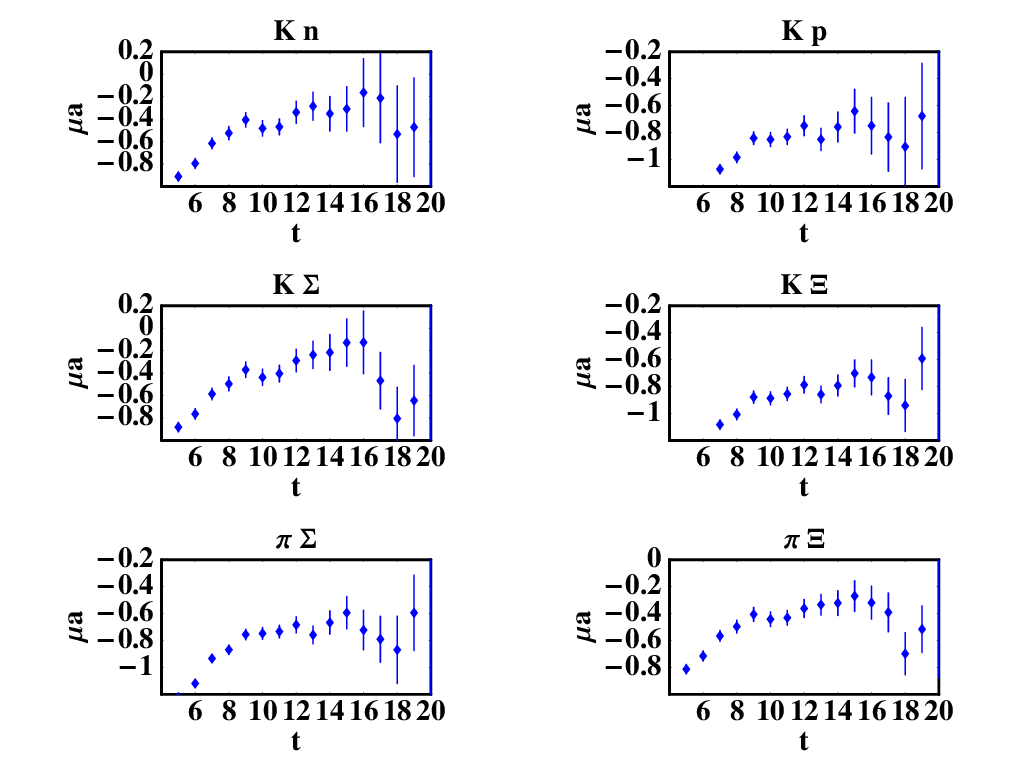}}
\caption{Effective scattering length (times reduced mass) plots for the six annihilation-free meson-baryon processes. For this
MILC ensemble the pion mass is roughly $600~{\rm MeV}$, $b\sim 0.125~{\rm fm}$ and $L\sim 2.5~{\rm fm}$.} \label{fig:MB}
\end{center}
\end{figure}
%%%%%%%%%%%%%%%%%%%%%%%%%%%%%%%%%%%%%%%%%%%%%%%%%%%%%%%%%%%%%%%%%%%%%%%%%%%%%

\section{Meson-Baryon Interactions}

\noindent Pion-nucleon scattering has long been considered a
paradigmatic process for the comparison of $\chi$-PT and experiment.
To this day, controversy surrounds determinations of the pion-nucleon
coupling constant and the pion-nucleon sigma term. While it would be
of great interest to calculate scattering parameters for this process
on the lattice, pion-nucleon correlation functions necessarily involve
disconnected diagrams. Indeed, considering the meson and baryon
octets, there are six processes that are free of annihilation:
$\pi^+\Sigma^+$, $\pi^+\Xi^0$, $K^+ p$, $K^+ n$, ${\bar K}^0
\Sigma^+$, and ${\bar K}^0 \Xi^0$. Preliminary lattice QCD results by
the NPLQCD collaboration now exist which use domain-wall valence
quarks on a rooted staggered sea~\cite{Torok}. An example of effective
scattering lengths is shown in Fig.~\ref{fig:MB}. An interesting
aspect of this system of six processes is that the $\chi$-PT
description at next-to-leading order contains two free
parameters~\cite{Liu:2006xja}. Hence the overconstrained nature of the
system provides an interesting test of chiral symmetry and the lattice
methodology.  It is also worth noting that an understanding of
meson-baryon energy levels is an essential ingredient in any attempt
to extract excited-baryon masses from lattice
calculations~\cite{Morningstar:2008mc}.

\section{Baryon-Baryon Interactions}

%%%%%%%%%%%%%%%%%%%%%%%%%%%%%%%%%%%%%%%%%%%%%%%%%%%%%%%%%%%%%%%%%%%%%%%%%%%%
\begin{figure}[hb!]
\begin{center}
\centerline{\includegraphics*[width=0.52\textwidth,clip,angle=0]{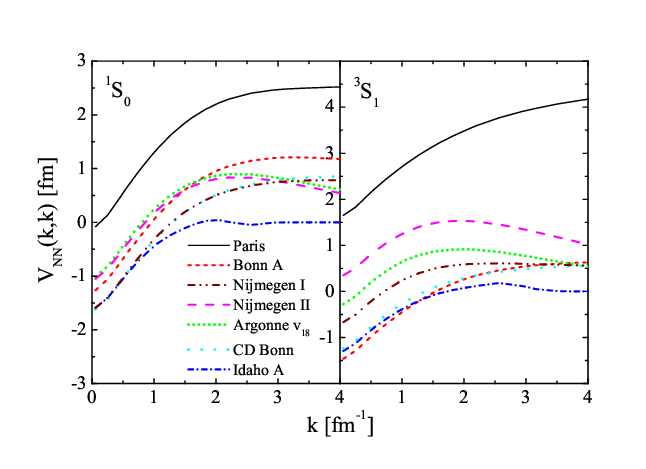}\includegraphics*[width=0.52\textwidth,angle=0]{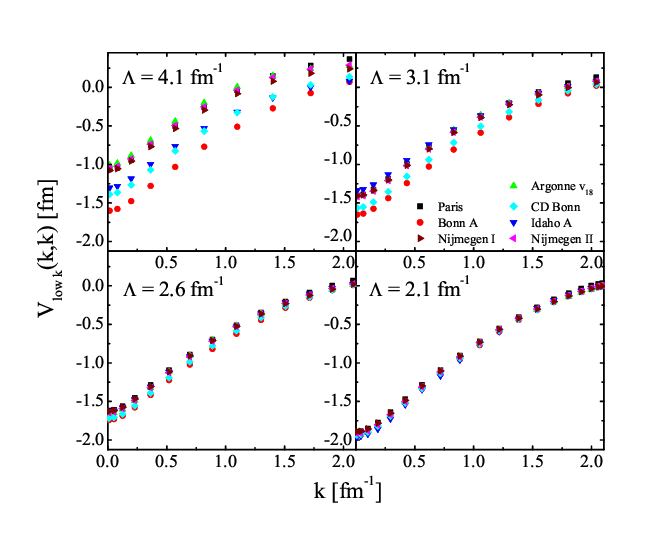}}
\caption{Left panel: momentum-space matrix elements for an assortment of bare NN potentials in the $\si$ and $\siii$ channels. Right panel: momentum-space matrix elements for NN potentials with short distance physics excluded beyond a cutoff $\Lambda$.} 
\label{fig:vlowk}
\end{center}
\end{figure}
%%%%%%%%%%%%%%%%%%%%%%%%%%%%%%%%%%%%%%%%%%%%%%%%%%%%%%%%%%%%%%%%%%%%%%%%%%%%%

\subsection{Potentials or Phase Shifts? A No-Go Theorem}

\noindent Modern nucleon-nucleon (NN) potentials fit the NN phase
shift ``data'' at low energies with a chi-squared of order one.  One
might then envisage calculating an NN potential directly from lattice
QCD which could be input into the Schr\"odinger equation to generate
first principles predictions for phase shifts. A priori, this seems
problematic; after all, unless the scattering particles are infinitely
heavy, the potential is not an observable in quantum mechanics, as a
unitary transformation will change the potential and the wavefunction
in such a way as to leave observables invariant. Therefore at best one
can calculate a potential that has been defined as a particular
lattice QCD correlation function with the understanding that this
choice is not unique.

With this defined NN potential in hand, could one compare it to modern
NN potentials? Modern NN potentials tend to treat the short-range part
of the NN force in completely different ways, indeed sometimes using
arbitrary parameterizations, while maintaining more or less the same
almost-perfect agreement with data at low energies. This is no
surprise if one thinks in the language of the renormalization
group. Ref.~\cite{Bogner:2003wn} has considered momentum-space matrix
elements of various modern NN potentials, $V_{NN}(k,k)$, that fit the
low-energy data and yet look quite different at short distances.  See
Fig.~\ref{fig:vlowk}. If one integrates out physics above a cutoff
$\Lambda$, one sees that as $\Lambda$ is reduced, the various
potentials, $V_{\rm low\,k}(k,k)$, approach a universal curve, thus
indicating that the details of the short-distance physics are
irrelevant to low-energy scattering data.  The bottom line is that if
one is able to compute the NN potential from QCD, then there is no
meaningful way in which the short-distance part of the potential may
be compared to phenomenological NN potentials. The only utility of the
potential would be to calculate phase shifts.

Recently, it has been claimed that the NN and
YN potentials can be extracted from the lattice
wavefunctions of two nucleons~\cite{Ishii:2006ec,Nemura:2008sp},
extending the technique that CP-PACS has successfully used to
determine $I=2$ $\pi\pi$ scattering parameters~\cite{Aoki:2005uf}.
Given the widespread attention that this work has received, it is
worth repeating here why this method is flawed~\cite{Detmold:2007wk}.

The NN correlation function measured on the lattice in Ref.~\cite{Ishii:2006ec} (IAH) is
\begin{eqnarray}
G_{NN}({\bf x},{\bf y},t) & = & \langle 0 | 
\hat {\cal O}_1({\bf x},t)_{\alpha}^{i} 
\hat {\cal O}_1({\bf y},t)_{\beta}^{j} 
\overline{J}(0) |0\rangle
\nonumber\\
& = & 
\sum_n\  \langle 0 | 
\hat {\cal O}_1({\bf x},0)_{\alpha}^{i} 
\hat {\cal O}_1({\bf y},0)_{\beta}^{j} 
|\psi_n\rangle
\langle\psi_n| \overline{J}(0) |0\rangle \ {e^{-E_n t}\over 2 E_n}
\ \ \ ,
\end{eqnarray}
where $\hat {\cal O}_1({\bf x},t)_\alpha^i$ is a nucleon interpolating
field with Dirac-index $i$, and isospin index $\alpha$. $\overline{J}$
is a wall-source on the initial time-slice $t_0=0$, and
$|\psi_n\rangle$ are the eigenstates of the Hamiltonian in the
finite-volume. In particular, $|\psi_n\rangle$ are states of definite
baryon number and isospin, and transform non-trivially under the hyper-cubic
group. Setting $\langle\psi_n| \overline{J}(t_0) |0\rangle =
A_n(t_0)$, at long times the correlation function becomes
\begin{eqnarray}
G_{NN}({\bf x},{\bf y},t) & \rightarrow &
 A_0(0)\ \langle 0 | 
\hat {\cal O}_1({\bf x},0)_{\alpha}^{i} 
\hat {\cal O}_1({\bf y},0)_{\beta}^{j} 
|\psi_0\rangle\ {e^{-E_0 t}\over 2 E_0}
\ \ \ ,
\end{eqnarray}
where $E_0$ is the ground-state energy shifted from $2M$ by
boundary effects ($\Delta E_2$ in the notation given above in eq.~\ref{eq:1}).
From this object, IAH generate the potential:
\begin{eqnarray}
U_{E_0}(r) & = & E_0 \ +\ {1\over 2 \mu} {\nabla^2 G_{NN}\over G_{NN}}
\ \ \ ,
\label{eq:IAHpot}
\end{eqnarray}
where $\mu$ is the reduced mass of the NN system. Here the energy
dependence of the potential $U_{E_0}(r)$ has been made explicit
and $\Psi=G_{NN}$ trivially satisfies the Schr\"odinger equation
for this potential.  IAH then assert that $\langle 0 |
\hat {\cal O}_1({\bf x},t_0)_{\alpha}^{i} \hat {\cal O}_1({\bf
  y},t_0)_{\beta}^{j} |\psi_0\rangle$ is proportional to
the non-relativistic, equal-time, Bethe-Salpeter (BS) wavefunction 
$\Phi^{ij}_{\alpha\beta}\equiv\langle 0 | N({\bf
  x},t_0)_{\alpha}^{i} N({\bf y},t_0)_{\beta}^{j} |\psi_0\rangle$,
where $N({\bf x},t)$ is a free-field nucleon annihilation operator.

%%%%%%%%%%%%%%%%%%%%%%%%%%%%%%%%%%%%%%%%%%%%%%%%%%%%%%%%%%%%%%%%%%%%%%%%%%%%
\begin{figure}[hb!]
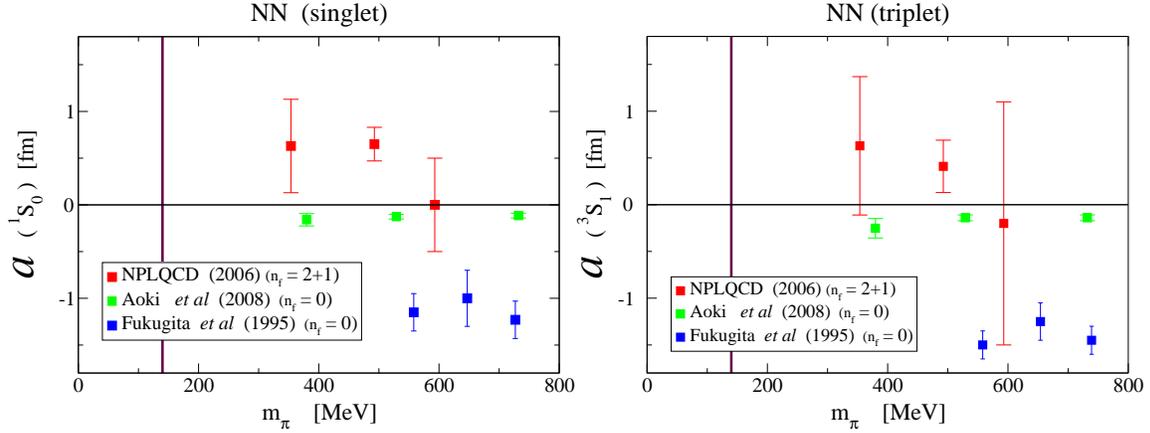

\begin{center}
\centerline{\includegraphics*[width=0.5\textwidth,clip,angle=0]{SING_NNlattice2008.eps}\includegraphics*[width=0.5\textwidth,angle=0]{TRIP_NNlattice2008.eps}}
\caption{The current status of lattice QCD calculations of the s-wave NN scattering lengths; left panel: $\si$; right panel: $\siii$.
The solid vertical line indicates the physical pion mass.} \label{fig:NN}
\end{center}
\end{figure}
%%%%%%%%%%%%%%%%%%%%%%%%%%%%%%%%%%%%%%%%%%%%%%%%%%%%%%%%%%%%%%%%%%%%%%%%%%%%%

However, this identification of the Bethe-Salpeter wavefunction is not correct
as the most general form for the matrix element is
\begin{eqnarray}
\langle 0 | \hat {\cal O}_1({\bf x},t_0)_{\alpha}^{i} 
\hat {\cal O}_1({\bf y},t_0)_{\beta}^{j} 
|\psi_0\rangle 
& = &  
Z_{NN}^{(S,I)}(|{\bf r}|)\ 
\langle 0 | N({\bf x},t_0)_{\alpha}^{i} N({\bf y},t_0)_{\beta}^{j}
|\psi_0\rangle
+\ldots
\, ,
\label{eq:full}
\end{eqnarray}
where $Z_{NN}^{(S,I)}(|{\bf r}|)$ is an unknown function that depends
on details of the composite sink, $\hat {\cal O}_{1,{\alpha}}^{i}\hat
{\cal O}_{1,{\beta}}^{j}$ and on the separation ${\bf r}={\bf x}-{\bf
y}$. The ellipses denote additional contributions from the tower of
states of the same global quantum numbers.

In the limit $|{\bf r}|\to\infty$, $Z_{NN}^{(S,I)}(|{\bf r}|)\to
(\sqrt{Z_N})^2$ (where $Z_N=|\langle 0 | \hat {\cal O}_{1}|
N\rangle|^2$) and the additional terms in eq.~(\ref{eq:full})
containing $p>2$ particles are suppressed.  Consequently the
scattering parameters can be rigorously extracted from
$G_{NN}$. However, inside the range of the NN interaction ($|{\bf
r}|<m_\pi^{-1}$), $G_{NN}$ depends explicitly on the interpolating
fields that are used.  Hence the ``potential'' defined in
eq.~\ref{eq:IAHpot} contains only a single piece of useful physics:
the phase shift $\delta(E_0)$, evaluated at the specific energy $E_0$,
which is precisely what one extracts using the finite-volume method
described above. The energy dependence of the BS equation has been
considered in toy models in Ref.~\cite{Aoki:2008yw}, however the
implications for the realistic problem are not clear.

Perhaps not surprisingly, this no-go theorem implies that the only
meaningful information about hadron-hadron scattering that can be
rigorously computed in lattice QCD consists of S-matrix elements.

\subsection{Current status of Baryon-Baryon Scattering}

%%%%%%%%%%%%%%%%%%%%%%%%%%%%%%%%%%%%%%%%%%%%%%%%%%%%%%%%%%%%%%%%%%%%%%%%%%%%
\begin{figure}[hb!]
\begin{center}
\centerline{\includegraphics*[width=0.5\textwidth,clip,angle=0]{NSigTRIP_493.eps}\includegraphics*[width=0.52\textwidth,angle=0]{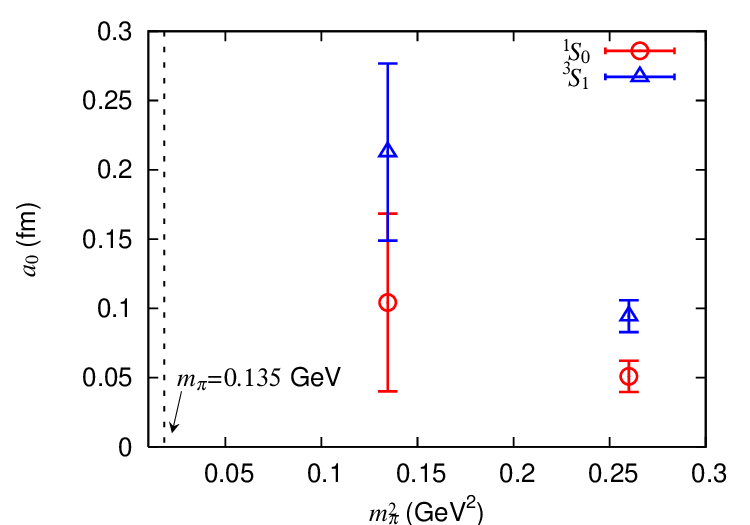}}
\caption{Lattice QCD calculations of the s-wave YN scattering lengths; left panel: the $n\Sigma^-$ $\siii$ phase shift evaluated
at center-of-mass momentum $|k|=261~{\rm MeV}$, compared to various potential models; right panel: the s-wave $p\Xi^0$ scattering lengths at various pion masses
in quenched QCD.} \label{fig:YN}
\end{center}
\end{figure}

\noindent Recently, the NPLQCD collaboration has performed the first
full-QCD calculation of the s-wave $NN$ scattering
lengths~\cite{Beane:2006mx}. At the pion masses used in these
calculations, the $NN$ scattering lengths were found to be of natural
size in both channels, and much smaller than the $L\sim 2.5~{\rm fm}$
lattice spatial extent. (See Fig.~\ref{fig:NN}, which also includes
the quenched calculations of
Refs.~\cite{Fukugita:1994ve,Aoki:2008hh}.)  The lowest pion mass
calculated ($\sim 350~{\rm MeV}$ ) is at the upper limit of where one
expects the EFT describing $NN$ interactions to be valid.  While the
NN system is clearly plagued by the signal to noise problem discussed
above, as the $NN$ signals improve with increased statistics, a
lattice QCD prediction of the low-energy scattering parameters will
become possible. However, it may well be the case that accurate
benchmarking for the NN system will initially be done with higher
partial waves, whose effective range parameters are of natural size
and dominated by pion physics.

Study of the interactions of hyperons with nucleons and nuclei is an
exciting area of nuclear physics, as mentioned in the introduction.
YN interactions influence the structure and energy-levels of
hypernuclei and are expected to be a basic input in studies of the
Equation of State of dense stellar matter. Initial investigations of
these interactions using lattice QCD have been carried through.  Here
scattering lengths of natural size are expected as there are probably
no YN bound states near threshold. As an example, Fig.~\ref{fig:YN}
(left panel) displays the spin-triplet $n\Sigma^-$ phase shift at
(center of mass) momentum $|k|=493~{\rm MeV}$ and $m_\pi\sim 350~{\rm
MeV}$ calculated by the NPLQCD collaboration in fully-dynamical
lattice QCD~\cite{Beane:2006gf}, compared to various potential models.
Fig.~\ref{fig:YN} (right panel) displays the s-wave $p\Xi^0$
scattering lengths in a recent quenched calculation, for various pion
masses~\cite{Nemura:2008sp}.

\section{Conclusion}

\noindent Lattice QCD calculations of two- and three-body interactions
of pions and kaons are now a precision science (for those channels
that do not involve disconnected diagrams). The study of multi-pion
systems has led to the first lattice QCD evidence of many-body
forces. While these results provide an important test of the basic
methodology for extracting many-body physics from lattice QCD, they
are also useful for the study of many-body physics like meson
condensation. It will be of great interest to see results of competing
calculations with different fermion discretizations in the meson sector.

A milestone for this area of research is to see a definitive signal 
for nuclear physics. Here one is plagued by a severe signal to noise
problem and, for the case of the NN interaction, a fine-tuned system that requires
a non-perturbative effective field theory description. However a great
deal of progress has been made in a short period; initial results
for NN and YN scattering parameters now exist in fully-dynamical
lattice calculations and the advent of petascale computing aligns
nicely with the need for very high-statistics calculations, which
promise to herald a golden age of exploration for nuclear physics.

\section*{Acknowledgments}

\noindent I thank W.~Detmold, T.C.~Luu, K.~Orginos, A.~Parre\~no,
M.J.~Savage, A.~Torok and A.~Walker-Loud for fruitful collaboration.
I would also like to thank P.~Lepage for a useful conversation.
This work was supported by NSF CAREER Grant No. PHY-0645570.

%\vfill\eject

\end{document}